\documentclass[a4paper,12pt]{article}
\usepackage{verbatim} 
\usepackage{jheppub} % for details on the use of the package, please
\usepackage[T1]{fontenc} % if needed
\usepackage{amsmath}
\usepackage[percent]{overpic}
\usepackage{wrapfig}
\usepackage{tabu}
\usepackage{diagbox}
\usepackage{slashbox}
\usepackage{pict2e}
\usepackage{graphicx}
\usepackage{tikz}
\usepackage{import}
\usepackage{accents}
\usepackage{mathrsfs,amsmath,amssymb,slashed}
\usepackage{slashed}

\usepackage{wrapfig}
\usepackage{tabu}
\usepackage{diagbox}
\usepackage{mathrsfs,amsmath,amssymb,amsthm,amsfonts,graphicx,hyperref,color}
\usetikzlibrary{decorations.pathmorphing}
\DeclareFontFamily{OT1}{rsfs}{}

\DeclareFontShape{OT1}{rsfs}{m}{n}{ <-7> rsfs5 <7-10> rsfs7 <10->rsfs10}{} 

\DeclareMathAlphabet{\mycal}{OT1}{rsfs}{m}{n}

\newcommand{\nn}{\nonumber}

\newcommand{\be}{\begin{equation}}
\newcommand{\ee}{\end{equation}}

\DeclareMathOperator{\extdm}{d}
\newcommand{\extd}{\extdm \!}

\newcommand{\dino}{\rho}
\newcommand{\dily}{x_0}
\newcommand{\dilx}{x_1}
 \title{\bf 
  {Holography in $\widehat{\text{CGHS}}$ Supergravity}}
\author[]{Hamid Afshar$^{\,a,b}$, Narges Aghamir$^{\,a}$}

\affiliation[a]{\it Department of Physics, Faculty of Science, Ferdowsi University of Mashhad, Mashhad, Iran}
\affiliation[b]{\it School of Physics, Institute for Research in Fundamental
 Sciences (IPM),\\ P.O.Box 19395-5531, Tehran, Iran} 

\emailAdd{ham.afshar@gmail.com, n.aghamir@gmail.com}

\abstract{
We study holographic aspects of 2D dilaton-supergravity in flat space-time using gauge theoretic BF formulation. The asymptotic symmetries in Bondi gauge and at finite temperature span a supersymmetric extension of the warped Virasoro algebra at level zero. The boundary action is determined such that the bulk variational principle is ensured and turns out to be a super-warped Schwarzian theory at the vanishing level. We also study the thermodynamics of the black hole saddle in this model.
}
\begin{document}
\maketitle
 \section{Introduction}
Holographic studies of dilaton-gravity theories in two dimensions (see \cite{Grumiller:2002nm} for a review on 2D dilaton-gravities) have been growing in recent years after the discovery of the duality between the Jackiw-Teitelboim (JT) gravity \cite{ Jackiw:1984je, Teitelboim:1983ux} and the low-temperature Sachdev-Ye-Kitaev (SYK) model \cite{Sachdev:1992fk,Sachdev:2010um,Kitaev:15ur,Kitaev:2017awl}. JT gravity provides a nearly AdS$_2$ description of the near-horizon geometry of nearly extremal black holes and the SYK model, on the other hand, is a nearly conformal interacting statistical quantum mechanical model of Majorana fermions with random couplings. The boundary dynamics of JT gravity under suitable boundary conditions are given in terms of a Schwarzian theory which emerges as the low energy effective action of the SYK model in the large $N$ (semi-classical) limit \cite{Kitaev:15ur,Sachdev:2010um,Kitaev:2017awl,Almheiri:2014cka, Maldacena:2016hyu, Maldacena:2016upp, Engelsoy:2016xyb, Cvetic:2016eiv}. This duality has been extended to the case of complex SYK model  \cite{Sachdev:2015efa,Davison:2016ngz,Bulycheva:2017uqj,Chaturvedi:2018uov,Gaikwad:2018dfc,Gu:2019jub} whose effective action improves to a warped Schwarzian theory \cite{Afshar:2019tvp}. It was shown in \cite{Afshar:2019axx} that the cSYK model in a certain double scaling limit is dual to a flat holographic bulk theory, originally denoted as $\widehat{\text{CGHS}}$ model \cite{Afshar:2019axx,Godet:2020xpk,Afshar:2020dth,Godet:2021cdl}. A natural plan is to investigate the supersymmetric version of this holographic picture in the semi-classical regime. The denomination $\widehat{\text{CGHS}}$ is due to the classical equivalence of this model with the conformally transformed matterless Callan–Giddings–Harvey–Strominger model \cite{Callan:1992rs}. The construction of the $\widehat{\text{CGHS}}$ model is based on the BF gauge theory formulation of Cangemi and Jackiw which was first introduced as a string-inspired model for gravity on a line \cite{Cangemi:1992bj}. This authenticates the name CJ gravity to this model as well \cite{Kar:2022vqy, Kar:2022sdc, Rosso:2022tsv}.\footnote{Throughout this paper, we sometimes refer to CJ as the classical model and to $\widehat{\text{CGHS}}$ as the quantum gravity model.} In contrast to the JT model,  boundary conditions in $\widehat{\text{CGHS}}$ (or CJ) model is ordered such that, it describes the near horizon geometry of the nearly non-extremal black holes (flat Rindler) \cite{Godet:2021cdl}.

 In this paper, we work on the bulk side where we have the $\widehat{\text{CGHS}}$ supergravity based on the BF-theory construction of Rivelles, Cangemi, and Leblanc \cite{Rivelles:1994xs, Cangemi:1993mj}.  The theory exhibits an infinite dimensional asymptotic symmetry which is the supersymmetric warped-Virasoro algebra with abelian Heisenberg subalgebra;
 \begin{equation}
 \begin{split}    
 \label{InftyAlgebra}
[L_n,L_m]&=(n-m)L_{n+m}+\frac{c}{12}(m^3-m)\delta_{m+n,0}\,,\\
[L_n,P_m]&=-mP_{n+m}-i\kappa \,(m^2+m)\delta_{m+n,0}\,,\qquad[P_m,P_n]=0\,,\\ 
[L_n,\Psi_r]&=\left(\frac{n}{2}-r\right)\Psi_{n+r}\,,\qquad\qquad[P_n,\Psi_r]=0\,,\\
\{\Psi_r,\Psi_s\}&=-\frac{1}{2\kappa}\sum_qP_{r+s-q}P_q+i(r+s-1)P_{r+s}-2\kappa(s^2-\frac{1}{4})\delta_{r+s,0}\,.
 \end{split}
 \end{equation}
 This extension of the Virasoro symmetry naturally accommodates for an extension of the corresponding Schwarzian Goldstone action which we systematically derive it using the bulk variational principle of the supergravity theory. The boundary action is then expected to be the effective action of a would-be supersymmetric SYK model in a scaling limit in the spirit of the statement in \cite{Afshar:2019axx} which is not addressed here. 

The structure of the paper is as follows, in section \ref{sec2} we introduce the bulk supergravity model. In section \ref{sec3} we develop the boundary conditions in the Bondi gauge at finite temperature and obtain the corresponding asymptotic symmetries. In section \ref{sec4} we develop the variational principle analysis and thereby obtain the boundary action together with integrability conditions. In section \ref{secA} we calculate the entropy of black hole solutions by imposing the holonomy condition on the solution space and evaluating both the on-shell action and also canonical charges. We also solve the stabilizer condition on our zero mode Euclidean background in this section.

 \par{Note added: While this work was in progress, ref. \cite{Rosso:2022tsv} was posted on the arXiv, whose results overlap
with some of ours.}
 
 \section{Supergravity model}\label{sec2}
The string-inspired dilaton-gravity CGHS model \cite{Callan:1992rs} after a dilaton-dependent Weyl rescaling can be formulated as a BF gauge theory \cite{Cangemi:1992bj, Afshar:2019axx}.  
The BF-theory formulation of the $\widehat{\text{CGHS}}$ supergravity model is based on the supersymmetric  extension of 2D Maxwell algebra (central extension of the 2D Poincar\'e algebra) \cite{Cangemi:1993mj,Rivelles:1994xs},
\begin{equation}\label{Lsuperalg}
\begin{split}
  [P_a, P_b]&=\epsilon_{ab}\tilde Z\,, \quad [J, P_a]={\epsilon_a}^bP_b\,,\\ [P_a, Q_\alpha]&=\frac{1}{2}(\Gamma_aU)_\alpha\,, \quad [J, Q_\alpha]=-\frac{1}{2}(\Gamma_3Q)_\alpha\,, \quad [J, U_\alpha]=-\frac{1}{2}(\Gamma_3U)_\alpha\,, \\ [K, Q_\alpha]&=-\frac{1}{2}(\Gamma_3U)_\alpha\,,\quad \left\{Q_\alpha, Q_\beta \right\}=({\Gamma^a})_{\alpha\beta}P_a-(\Gamma^3)_{\alpha\beta}K\,, \quad \left\{Q_\alpha, U_\beta \right\}=-(\Gamma^3)_{\alpha\beta}\tilde Z\,,
\end{split}
\end{equation}%\tcr{It is better to derive the algebra as the expansion of the AdS-algebra}.
where $a,b=0,1$ specify Lorentz indices while the Greek indices $\alpha,\beta=1,2$ run over the components of two dimensional spinors. The two-dimensional gamma matrices in \eqref{Lsuperalg} are given as
\begin{align}
({\Gamma_0})_\alpha{}^\beta=\begin{pmatrix}
0 & -1  \\
1 & 0  \\
\end{pmatrix},\quad ({\Gamma_1})_\alpha{}^\beta= \begin{pmatrix}
0 & 1  \\
1 & 0  \\
\end{pmatrix}, \qquad
 (\Gamma_3)_\alpha{}^\beta=-(\Gamma_0\Gamma_1)_\alpha{}^\beta= \begin{pmatrix}
1 & 0  \\
0 & -1  \\
\end{pmatrix}\,,
\end{align}
and $\Gamma_{\alpha\beta}={{\left ( \Gamma \right )}_\alpha}^\gamma C_{\gamma \beta}=(\Gamma C^{-1})_{\alpha\beta}$ where  $C=\Gamma^0=-\Gamma_0$ is the charge conjugation matrix that allows us to raise and lower the spinor indices using $C^{\alpha \beta}$ and $C_{\alpha \beta}$ as components of the matrix $C^T$ and $C^{-1}$  respectively
\cite{freedman2012supergravity,Murayama2007}. 

This algebra can be derived as the extension of the $N=1$ 2D super-Poincar\'e algebra which itself is a contraction of the osp(2,1) algebra ($N=1$ supersymmetric AdS$_2$ algebra). The extension is based on introducing the  
contraction parameter $\sigma=1/\ell^2$ and expanding osp(2,1) generators appropriately in the contraction parameter i.e. $J^{(0)}\equiv J$, $J^{(1/2)}\equiv \sqrt\sigma K$, $J^{(1)}\equiv \sigma \tilde Z$ and $Q_\alpha^{(0)}\equiv Q_\alpha$, $Q_\alpha^{(1/2)}\equiv \sqrt\sigma U_\alpha$. This leads us to the 2D super Maxwell algebra \eqref{Lsuperalg} with extra fermionic generators $U_\alpha$, which are nilpotent and their anti-commutators vanish. So one may call it $N=2$ supersymmetry. 

The superalgebra \eqref{Lsuperalg} admits quadratic  and linear Casimir elements as follows
\begin{align}
   \mathbf{C}_2&=P_aP^a+K^2+J\tilde Z+\tilde ZJ+\frac{1}{2}C^{\alpha\beta}(Q_\alpha U_\beta+U_{\beta} Q_{\alpha})\,,\\
   \mathbf{C}_1&=\tilde Z\,.
\end{align}
Correspondingly it has the following graded invariant non-degenerate bilinear form;
\begin{align}\label{bilinearform}
    \langle P_a,P_b\rangle=\eta_{ab},\quad \langle J,\tilde Z \rangle=1,\quad \langle J,J \rangle=  b,\quad \langle K,K \rangle=1,\quad \langle Q_\alpha,U_\beta \rangle=2C_{\alpha\beta}\,,
\end{align}
where $\eta=\text{diag}(-1,+1)$ and $ b$ is an arbitrary constant. The emphasis on the presence of $b$ in the bosonic bilinear form goes back to \cite{Nappi:1993ie} while in the present context, it was first introduced in \cite{Afshar:2020dth} where the theory with $b\neq0$ was called $\mathit{twisted}$ $\widehat{\text{CGHS}}$ model.\footnote{It was shown in \cite{Afshar:2020dth} that the presence of $b$ leads to a Schwarzian derivative term in the boundary Lagrangian, see also section \ref{sec4}.}

Since this superalgebra admits an invariant bilinear form one can construct its BF gauge theory as a 2D dilaton-supergravity,
\begin{align}
    I_0=\kappa\int\langle B,F\rangle\,,\label{2.2}    
\end{align}%+I_{\tiny{\text{bndry}}}
where $B$ is the algebra valued scalar field and $F$ is the 2-form field strength associated to the gauge field {$A$} of the algebra \eqref{Lsuperalg}. It is easy to check that this theory is invariant under the following general gauge transformation (with parameter $\Lambda$) associated with these fields 
\begin{align}
    \delta A= \extd\Lambda+ \left [A, \Lambda  \right ]\,,\qquad\delta B=\left [ B, \Lambda \right ]\,.\nn
\end{align}
By choosing $A$ and $B$ as
\begin{equation}\label{AAAAAA}
\begin{split}
      A=e^aP_a+\omega J+ a\,\tilde Z+\psi^\alpha Q_\alpha+\chi^\alpha U_\alpha+vK\,,\\ B=X^aP_a+Y J+X\tilde Z+\phi^\alpha Q_\alpha+\rho^\alpha U_\alpha+WK\,,
      \end{split}
\end{equation}
we reproduce the BF theory supergravity action in the first-order formulation \cite{Cangemi:1993mj, Rivelles:1994xs}
\begin{align}
    I_0&=\kappa\int X_a\left (\extd e^a+{\epsilon_b}^a\omega e^b-\frac{1}{2}\bar{\psi}\Gamma^a\psi \right )+Y\left (  b\extd\omega+\extd a+\frac12\epsilon_{ab} e^ae^b+\bar{\psi}\Gamma_3\chi\right )+X\extd\omega\nn\\&\qquad\quad+2\bar{\rho} \mathcal D\psi+\bar{\phi}\left ( 2\mathcal D\chi+e^a\Gamma_a\psi-\frac12v\Gamma_3\psi-\frac12W\Gamma_3\psi\right )+W(\extd v+\frac{1}{2}\bar{\psi}\Gamma_3\psi)\,,\
\end{align}
where $\mathcal D\psi=\extd\psi-\frac12 \omega \Gamma_3\psi$, and $\bar\lambda\equiv\lambda^TC$ (leading to $\bar{\lambda}\Gamma\psi=\lambda^\alpha\Gamma_\alpha{}^\beta\psi_\beta$ for a typical matrix $\Gamma$). The field content of this theory is as follows: $X^a$ is a pair of Lagrange multipliers that require the torsion constraint, $X$ is the dilaton field, $Y$ is a scalar field that is constant on-shell required by the field equation of the gauge field $a$. The $\phi^\alpha, \rho^\alpha$ stand for the dilatino spinor fields, and $\psi^\alpha, \chi^\alpha$ are the gravitino spinor vector fields. $e^a$ corresponds to the Zweibein vector field and $\omega$ is the dualized spin connection. The gauge field $v$ and the scalar field $W$ are $R$-symmetry fields.  This action is invariant under the local supersymmetry transformation $\Lambda=\epsilon^\alpha Q_\alpha+\zeta^\alpha U_\alpha$,
\begin{equation}
\begin{split}
&\delta e^a=-\bar{\psi}\Gamma^a\epsilon,\qquad \delta \omega=0,\qquad \delta X^a=-\bar{\phi}\Gamma^a\epsilon,\qquad \delta X=\bar{\phi}\Gamma_3\zeta+\bar{\rho}\Gamma_3\epsilon,\qquad \delta a =\bar{\psi}\Gamma_3\zeta+\bar{\chi}\Gamma_3\epsilon,\\& \delta \chi^\alpha=\mathcal D\zeta^\alpha+\frac12 e^a(\bar{\zeta}\Gamma_a)^\alpha-\frac12v(\bar{\epsilon}\Gamma_3)^\alpha,\qquad \delta \psi^\alpha=\mathcal D{\epsilon}^\alpha,\qquad \delta v=\bar{\psi}\Gamma_3\epsilon,\qquad \delta Y=0,\\&\delta\rho^\alpha=\frac12\left ( X^a(\bar{\epsilon} \Gamma_a)^\alpha-Y(\bar{\zeta} \Gamma_a)^\alpha-W(\bar{\epsilon} \Gamma_3)^\alpha\right ),\qquad \delta W=\bar{\phi}\Gamma_3\epsilon,\qquad \delta \phi^\alpha=-\frac12Y(\Bar{\epsilon}\Gamma_3)^\alpha\,.
\end{split}
\end{equation}
\subsection*{Light-cone algebra}
We work in the component form of the Lorentzian superalgebra \eqref{Lsuperalg}, whose light-cone form with $P_\pm=P_1\pm P_0$ is \footnote{
% In the Euclidean algebra, we use the symmetric metric $\delta_{ab}$  with $a, b = 1, 2$ and  we do not distinguish between the up and down flat indices. 
The closed form of the Euclidean superalgebra, with all generators being Euclidean, where we do not distinguish between the up and down flat indices has the same form as the Lorentzian one with $\epsilon_{21}=1$ in \eqref{Lsuperalg} after introducing the following analytic continuation in gamma matrices 
\begin{align}
    \Gamma_2^{\text{\tiny E}}=-i{\Gamma_0}, \qquad \Gamma_1^{\text{\tiny E}}= {\Gamma_1}\,,\qquad \Gamma_3^{\text{\tiny E}}= {\Gamma_3}\,.
\end{align}
The spinor indices in the Euclidean case are raised and lowered with the Euclidean charge conjugation  matrix $C^{\text{\tiny E}}=-\Gamma_2^{\text{\tiny E}}$ \cite{analytic}. Alternatively, one may start with \eqref{superalg} and use the following analytic continuation;
\begin{align}
    P_2^{\text{\tiny E}}=-iP_0\,,\qquad P_1^{\text{\tiny E}}=P_1\,,\qquad J_{\text{\tiny E}}=iJ\,,\qquad Z_{\text{\tiny E}}=iZ\,.
\end{align}
}
%($P_\pm=P_1\pm P_0$)
\begin{align}\label{superalg}
[P_+,P_-]&=Z\,, \quad [J,P_+]=-P_+\,,\quad [J,P_{-}]=P_-\,,\nn \\ 
   [P_-,Q_1]&=U_2\,,\quad[P_+,Q_2]=U_1\,,\quad [P_+,Q_1]=0=[P_-,Q_2]\,,\quad[J,Q_{1,2}]=\mp\frac{1}{2}Q_{1,2}\,, \nn\\ [J,U_{1,2}]&=\mp\frac{1}{2}U_{1,2}\,, \quad [K,Q_{1,2}]=\mp\frac{1}{2}U_{1,2}\,,\quad \left\{Q_1,Q_1 \right\}=P_+\,, \quad \left\{Q_2,Q_2 \right\}=-P_-\,,\nn \\ \left\{Q_1,Q_2 \right\}&=K,\quad \left\{Q_1,U_1 \right\}=0=\left\{Q_2,U_2 \right\}\,, \quad \left\{Q_1,U_2 \right\}=- \frac{Z}{2}=\left\{Q_2,U_1 \right\}\,,%\label{2.9}
\end{align}
 where we set $Z=-2\tilde Z$.
 The corresponding bilinear form \eqref{bilinearform} after an overall $1/2$ rescaling takes the following form in the light-cone 
\begin{align}\label{bilinearcomp}
 \langle P_+,P_-\rangle=-\left<J,Z \right>=2\left<K,K \right>=1,\quad \left<J,J \right>=b,\quad \left<Q_1,U_2 \right>=-\left<Q_2,U_1 \right>=-1\,.
\end{align}
Throughout the paper, we use the algebra \eqref{superalg} and the bilinear form \eqref{bilinearcomp}.

\section{Asymptotic symmetries}\label{sec3}
 In this section, we explore the question, what are the symmetries that act on the phase space of the theory at finite temperature? 
We confine the phase space of the theory by adopting the following boundary conditions on the gauge field  
 \begin{align}\label{bndrcd}
  A_u&=\mathcal{T}(u)P_++P_-+\mathcal{P}(u)J+\psi(u) Q_1\,,   \qquad A_r=0\,.
 \end{align}
 The set of field configurations \eqref{bndrcd} defines our phase space. Of course, the full phase space is determined by adding the on-shell configuration for the $B$-field. However, as we shall see for the purpose of this section it is enough to focus on \eqref{bndrcd}. We postpone the discussion on the $B$-field to the next section. 
Boundary conditions \eqref{bndrcd} effectively define the theory within the truncated algebra where the generators $U_1$, $Q_2$, and $K$ are turned off.  The bosonic part of the boundary condition \eqref{bndrcd} was introduced in \cite{Afshar:2019axx} which reproduces the flat Rindler-type black hole in the metric formulation in the Bondi-gauge, see section \ref{secA}. The  fermionic extension on this background in \eqref{bndrcd} along $Q_1$ is introduced from the fact that in the superalgebra, this generator has the only non-trivial commutator with $P_-$, resulting in a linear equation to solve as demanded by the Hamiltonian reduction. This consistent set of boundary conditions are preserved $\delta_\epsilon A=\mathcal{O}(\delta A)$ by the following gauge transformation.  
 \begin{align}\label{gaugetrans}
     \epsilon=\epsilon^+P_++\varepsilon P_-+\epsilon^JJ+\sigma Z+\epsilon^Q Q_1+\chi U_2
 \end{align}
 where
 \begin{subequations}\label{epsJQ}
      \begin{align}
     \epsilon^+&=\mathcal{T}\varepsilon+\sigma'-\frac{1}{2}\psi \chi, \,\,\\
     \epsilon^J&=\mathcal{P} \varepsilon+{\varepsilon}'\,,\\
     \epsilon^Q&=\psi \varepsilon-\frac{1}{2}\mathcal{P} \chi-{\chi}'\,.
 \end{align}
 
 \end{subequations}
 The following gauge transformation is then induced on the state-dependent functions
\begin{subequations}\label{variations}
\begin{align}
     \delta \mathcal{T}&={\epsilon^+}'-\mathcal{P} \epsilon^++\mathcal{T}\epsilon^J+\psi \epsilon^Q=\mathcal{T}'\varepsilon+2\mathcal{T}\varepsilon'-\mathcal{P}\sigma'-\frac32\psi\chi'-\frac12\psi'\chi+\sigma''\,,\\
     \delta \mathcal{P}&={\epsilon^J}'=(\mathcal{P} \varepsilon+{\varepsilon}')'\,,\\
     \delta \psi&={\epsilon^Q}'-\frac{1}{2}(\mathcal{P} \epsilon^Q-\psi\epsilon^J)=\frac32\psi\varepsilon'+\psi'\varepsilon+\frac14\Big(\mathcal{P}^2-2\mathcal{P}'\Big)\chi-\chi''\,.
 \end{align}
 \end{subequations}
The transformation rule \eqref{variations} determine the conformal weights of each field as $2$ for $\mathcal{T}$, $1$ for $\mathcal{P}$ and $3/2$ for $\psi$.  The twisted Sugawara combination $\mathcal M=\mathcal P^2-2\mathcal P'$ plays a role similar to the stress tensor with conformal weight $2$.
\subsection*{Euclidean theory}
We consider the theory at finite temperature $T=1/\beta$ by performing a Wick rotation $u\to iu\equiv \tau$ and working in Euclidean  periodic Bondi time $\tau\sim \tau + \beta$.
 Since  the theory is defined 
 % \comment{is defined} 
 at  finite temperature, we can introduce varied generators of symmetries acting on the phase space using the invariant pairing between the parameter of gauge transformation and the fluctuations  of the connection \cite{Afshar:2015wjm, Afshar:2020dth}
\begin{align}\label{VarGen}
\delta\mathcal{C}=\kappa\oint\langle\delta A_\tau,\epsilon\rangle\,,
\end{align}
where $A_\tau=-iA_u$ and the circle integration is defined as $\oint =\int_0^{\beta}\extd \tau$. 
% \tcr{Here we assume that the coordinate $u$ is periodic. Our phase space \cite{Afshar:2015wjm} suggests this choice.} 
These generators are defined by using the invariant bilinear form of the theory \eqref{bilinearcomp}. Due to the Euclidean analytic continuation,  we should replace $\partial_u$ in all equations \eqref{variations} with $i\partial_\tau$, however, these equations remain unchanged with $' \equiv\partial_\tau$ once we also Wick rotate variables due to their conformal weights,
\begin{align}\label{WickRot}
    \mathcal P\to i\mathcal P\,,\qquad \mathcal T\to -\mathcal T\,,\qquad\psi\to i^{3/2}\psi\,,\qquad
    \varepsilon\to -i\varepsilon\,,\qquad\sigma\to\sigma\,,\qquad\chi\to i^{-1/2}\chi\,.
\end{align}
We can split the varied generators \eqref{VarGen} according to the parameters of gauge transformation
\begin{align}
\delta \mathcal{C}=\delta\mathcal{C}[\varepsilon]+\delta\mathcal{C}[\sigma]+\delta\mathcal{C}[\chi]
\end{align}
where after applying the Wick rotation \eqref{WickRot} we get
\begin{align}
\delta\mathcal C[\varepsilon]=\kappa\oint(\delta\mathcal T\,\varepsilon+b\,\delta\mathcal P\epsilon^J)\,,\quad
\delta\mathcal C[\sigma]=-\kappa\oint\delta\mathcal P\sigma\,,\quad
\delta\mathcal C[\chi]=-\kappa\oint\delta\psi\,\chi\,.
\end{align}These pairings are integrable and coincide with pairing in the coadjoint orbit method between adjoint and coadjoint vectors. We then introduce the Fourier-mode generators 
 \begin{subequations}\label{generators}
 \begin{align}
L_n&=\mathcal C[\varepsilon=e^{\frac{2\pi}{\beta}in\tau}]=\kappa\oint\left(\mathcal T+b\,\Big(\frac12\mathcal P^2-\mathcal{P}'\Big)\right)e^{\frac{2\pi}{\beta}in\tau}\,,\\
P_n&=\mathcal C[\sigma=e^{\frac{2\pi}{\beta}in\tau}]=-\kappa\oint \mathcal{P}\,e^{\frac{2\pi}{\beta}in\tau}\,,\\
\Psi_r&=\mathcal C[\chi=e^{\frac{2\pi}{\beta}ir\tau}]=-\kappa\oint \psi\,e^{\frac{2\pi}{\beta}ir\tau}\,,
\end{align}
 \end{subequations}
with $n\in\mathbb{Z}$ and $r\in\mathbb{Z}/2$ if the fermionic sector fulfills anti-periodic (Neveu-Schwarz) boundary conditions and $r\in\mathbb{Z}$ for periodic (Ramond) boundary conditions.

We can use the definition of Poisson brackets $\delta_2\mathcal C_1=[\mathcal C_1,\mathcal C_2]$  where upon inserting the variations \eqref{variations} into \eqref{generators} leads to the commutators and anti-commutators
\begin{subequations}\label{InftyAlg}
\begin{align}
[L_n,L_m]&=(n-m)L_{n+m}+b\kappa\,m^3\delta_{m+n,0}\,,\label{InftyAlg1}\\
[L_n,P_m]&=-mP_{n+m}-i\kappa \,m^2\delta_{m+n,0}\label{InftyAlg2}\,,\\ 
[L_n,\Psi_r]&=\left(\frac{n}{2}-r\right)\Psi_{n+r}\,,\\
\{\Psi_r,\Psi_s\}&=M_{r+s}-2\kappa\, s^2\delta_{r+s,0}\label{InftyAlg3}\,,
\end{align}
\end{subequations}
where $M_n=-\frac{1}{2\kappa}\sum_qP_{n-q}P_q+inP_n$ and some appropriate rescalings of the generators have been implemented. One can check that the Jacobi identities are all satisfied. Especially the non-trivial Jacobi identity among $(L_m,\Psi_r,\Psi_s)$ holds,
\begin{align}
    \left\{ \Psi_r,\left [\Psi_s,L_m  \right ]\right\}&- \left\{ \Psi_s,\left [L_m ,\Psi_r \right ]\right\}+\left [ L_m,\left\{\Psi_r,\Psi_s \right\} \right ]=0\,.
\end{align}
In the infinite dimensional algebra \eqref{InftyAlg} if we shift the zero modes as $L_0\to L_0+b\kappa/2$ and $P_0\to P_0+i\kappa$, the  last terms on the right hand side of (anti-)commutators in \eqref{InftyAlg1}, \eqref{InftyAlg2} and \eqref{InftyAlg3}  change such that we get the supersymmetry algebra \eqref{InftyAlgebra} with $c=12b\kappa$.
The wedge subalgebra of \eqref{InftyAlgebra} is then spanned by the subset $\{P_0$, $P_{-1}$, $L_0$, $L_{1}$, $\Psi_{\pm1/2}\}$ with the following non-zero (anti-)commutators;\footnote{We could alternatively choose the subalgebra to be spanned by $\{P_0$, $P_{1}$, $L_0$, $L_{-1},\Psi_{\pm1/2}\}$. In this case one needs to do the shift in \eqref{InftyAlg} as $ P_0\to P_0-i\kappa$.}
\begin{align}\label{wedge1}
    [L_0,L_{1}]&=-L_{1}\,,\qquad [L_{1},P_{-1}]=P_0\,,\qquad[L_0,P_{-1}]=P_{-1}\,,\nn\\
    [L_0,\Psi_{\pm 1/2}]&=\mp\frac12\Psi_{\pm 1/2}\,,\qquad\quad [L_{1},\Psi_{-1/2}]=\Psi_{1/2}\,,\\\qquad\{\Psi_{1/2},\Psi_{-1/2}\}&=-iP_0+\mathcal{O}(1/\kappa)\,,\qquad\{\Psi_{-1/2},\Psi_{-1/2}\}=-2iP_{-1}+\mathcal{O}(1/\kappa)\,.\nn
\end{align}
In the semi-classical large-$\kappa$ limit we can then pick the vacuum of the theory by demanding that it is invariant under this subalgebra 
which corresponds to the six-dimensional super-Maxwell algebra with the following identification;
\begin{align}
    & L_{1} = P_-\quad   P_0=- Z,\quad L_0=-J, \quad  P_{-1}= P_+,\quad \Psi_{1/2}=\sqrt{-2i} U_2,\quad \Psi_{-1/2}=\sqrt{-2i} Q_1\,.
\end{align}
Alternatively one may consider the following six-dimensional subalgebra of the infinite-dimensional algebra \eqref{InftyAlgebra} which is spanned by the subset $\left\{ P_0,  P_{\pm1}, L_0, \Psi_{\pm1/2}\right\}$, as
\begin{align}\label{wedge2}
    [L_0,P_{1}]&=-P_{1}\,,\qquad[L_0,P_{-1}]=P_{-1}\,,\nn\\
    [L_0,\Psi_{\pm 1/2}]&=\mp\frac12\Psi_{\pm 1/2}\,,\qquad \{\Psi_{-1/2},\Psi_{-1/2}\}=-2iP_{-1}\,,\\\qquad\{\Psi_{1/2},\Psi_{-1/2}\}&=-iP_0\,,\qquad\{\Psi_{1/2},\Psi_{1/2}\}=-iP_{1}\,.\nn  
\end{align}
The superalgebra \eqref{wedge2} is isomorphic to the $\mathcal N=(1,1)$ 2D super-Poincar\'e algebra with $L_0$ as the boost generator and $P_{\pm}$ as translation generators, $\psi_{\pm1/2}$ as Majorana-Weyl supercharges and $P_0$ as a real central term. 
Wedge subalgebras \eqref{wedge1} and \eqref{wedge2} essentially define two distinct {\it flat} vacua in a theory whose ultimate asymptotic symmetries are the same as \eqref{InftyAlgebra}.  
It would be interesting to explore the quantum theory defined by the Minkowski vacuum \eqref{wedge2}\footnote{For 2D dilaton SUGRA based on $\mathcal N=(1,1)$ supersymmetry we refer to \cite{Bergamin:2004us}.}, though our focus here is on the first supersymmetric vacuum \eqref{wedge1} which has the interpretation of a 2D supersymmetric Rindler space-time  compatible with our boundary conditions \eqref{bndrcd}. 

 In addition to these two flat vacua, the infinite-dimensional algebra \eqref{InftyAlgebra} is also compatible with the AdS$_2$ vacuum. In this case its subalgebra is spanned by the subset $\left\{ P_0,  L_{\pm1}, L_0, \Psi_{\pm1/2}\right\}$, 
\begin{align}\label{wedge3}
  [L_0,L_{1}]&=-L_{1}\,,\qquad [L_{1},L_{-1}]=2L_0\,,\qquad[L_0,L_{-1}]=L_{-1}\,,\nn\\
    [L_0,\Psi_{\pm 1/2}]&=\mp\frac12\Psi_{\pm 1/2}\,,\qquad\quad [L_{1},\Psi_{-1/2}]=\Psi_{1/2}\,,\\\qquad\{\Psi_{1/2},\Psi_{-1/2}\}&=-iP_0\,,\qquad\{L_{-1},\Psi_{1/2}\}=-\Psi_{-\frac12}\,.\nn  
\end{align}
The bosonic part of this superalgebra is SL(2,$\mathbb R$)$\times$U(1). 

\section{Variational principle}\label{sec4}
The action principle tells that the physical equations of motion must extremize the action. This implies that our action should be differentiable (at least on-shell).
 A generic variation of the BF bulk action \eqref{2.2} when equations of motions hold, equals to a boundary term
 \begin{align}\label{genericVar}
   \delta I_0\approx\kappa\oint\langle B,\delta A_\tau\rangle\,,
\end{align}
where  $\approx$ denotes on-shell equality.
This variation is generically non-zero and thus the action is not differentiable at on-shell configurations and thus the variational principle is ill-defined. 
 In order to make the theory well-defined we need to define our action functional up to an appropriate boundary term $I_\partial=\kappa\oint L$ such that the improved action is differentiable,  
 \begin{align}\label{totalaction}
     I[B,A]=I_0+\kappa\oint L\,.
 \end{align}
 In general, the fields $B$ and $A$ are independent off-shell. However, we can relate them using the field equation $\extd B +[A, B]=0$ such that the varied action $\kappa\oint\delta L$  cancels the boundary term \eqref{genericVar}. Our method for finding the boundary Lagrangian $L$ is to use half of the field equations to write the boundary integrand in \eqref{genericVar} as a total integrable quantity $V=-L$ up to possible extra variation terms $\delta V_i$; 
 \begin{align}\label{Varpatern}
     \langle B,\delta A_\tau\rangle \approx \delta V+ C_i\delta V_i+\cdots
 \end{align}
where $\cdots$ are total derivative terms  on-shell and thus  can be thrown away. 
The final variation of the action \eqref{totalaction} is thus
 \begin{align}
\delta I \approx \kappa\oint C_i\delta V_i\,.
\end{align}
Now if we use all field equations to show that $C_i$'s are indeed constants of motions, they can be drawn out of the integral and we are left with some integrability conditions $\delta\oint V_i=0$ which should be solved and finally, we have $\delta I\approx0$. In the following, we apply this method first to the case of JT supergravity as an example and then to the supersymmetric (twisted-) $\widehat{\text{CGHS}}$ model (or Cangemi-Jackiw supergravity) which is the focus of our work.

 \subsection{Jackiw-Teitelboim supergravity}

The JT supergravity has a BF-theory description in terms of the osp(2,1) superalgebra. The details of the derivation of the (super-)Schwarzian boundary term can be found in \cite{Cardenas:2018krd} based on superfield formalism. Here in this section, we rederive the boundary action using our method described above. Our starting point is the variation of the action which in this case is given by
\begin{align}\label{VariationactionJT}
    \delta I_0^{\tiny{\text{\tiny {JT}}}}=\kappa\oint\langle B,\delta A_\tau\rangle=\kappa\oint\left(x\delta  \mathcal{L}-2\rho\delta\psi\right)
\end{align}
where ($\psi,\mathcal{L}$) and ($x,\rho$) are boundary values of the gauge field $A$ and the scalar field $B$ correspondingly. The following equations relate these two sets
\begin{align}
&\mathcal{L}x^2+\frac{1}{2}xx''-\frac{1}{4}(x')^2+3x\psi\rho+2\rho\rho'-{C}=0\,,\label{Casimireq}\\
& \frac{3}{2}\psi x'+\psi'x-\mathcal{L}\rho-\rho''=0\,,
\end{align}
where $C$ is a constant. With some manipulation, we can solve the second-order differential equation of $\psi$
in terms of a Grassman variable $\eta$ as,
\begin{align}\label{spinorCas}%\label{rhoetaJT}
        \psi x^{\frac{3}{2}}-x\left(\frac{\rho}{\sqrt{x}}\right)'- C \eta x^{-\frac{1}{2}}=\text{const.}\,,\qquad\text{with}\qquad  \left(\frac{\eta}{\sqrt{x}}\right)'=\rho x^{-\frac{3}{2}}\,.
\end{align}
The $x$-dependence in the denominator of the change of variable between $\rho$ to $\eta$ is such that the conformal weight is preserved. This choice also matches with the superconformal transformation encountered in the superfield formalism \cite{Cardenas:2018krd, Fan:2021wsb}. As we shall see later this choice also works for the case of $\widehat{\text{CGHS}}$ supergravity.

We can now use the field equations \eqref{Casimireq} and \eqref{spinorCas} for substituting and integrating the varied boundary action \eqref{VariationactionJT},
\begin{align}
    x\delta \mathcal{L}&=
        \delta\frac{{C}}{x}+{C}\delta\frac{1}{x}-\frac{1}{2}\left (x\delta(x'/x)\right )'-\delta\left ( 3\psi\rho+2\frac{\rho\rho'}{x} \right )+\delta x\left ( 3\frac{\psi\rho}{x}+2\frac{\rho\rho'}{x^2} \right )\,,\\
 \rho\delta\psi&=\frac12\delta x\left ( 3\frac{\psi\rho}{x}+2\frac{\rho\rho'}{x^2} \right )+\frac12\delta\frac{\rho\rho'}{x}+\delta\left(\frac{ C \rho\eta}{x^2}\right) + \frac12 C\delta\left(\frac{\eta\eta'}{x}\right)+\cdots\,,
\end{align}
where in the last term of the second equality we used the relation \eqref{spinorCas} and $\cdots$ are total derivative terms which we ignore. Finally, we have
\begin{align}\label{bondryactionJT}
     x\delta \mathcal{L}-2\rho\delta\psi&=\delta\left(\frac{ C}{x}(1+2\eta\eta')-3(\psi\rho+\frac{\rho\rho'}{x})\right)+ C\delta\left(\frac1x(1-\eta\eta')\right)\,.
\end{align}
This shows that the Lagrangian density is integrable providing that 
\begin{align}\label{JTintegrabcond}
    \delta\oint \frac1x(1-\eta\eta')=0\,.
\end{align}
We can identify the integrand as
\begin{align}\label{diffeos}
    f'=\frac{\hat x}{x}(1-\eta\eta')\,,
\end{align}
where the constant $\hat x$ is a fixed quantity ($\delta \hat x=0$) and $f'$ is a quasiperiodic function, $\oint f'=\beta$. The boundary action is thus
identified as minus the integral of the total variation term in \eqref{bondryactionJT}. After some algebra, we can write the boundary action in terms of $(f,\eta)$ upon using the relation \eqref{spinorCas} and \eqref{diffeos},
\begin{align}
  I_\partial^{\text{\tiny{JT}}}[f,\eta]&=-\kappa\oint\Big[(1+2\eta\eta')\mathcal L x-\frac14(1+\eta\eta')\frac{x'^2}{x}-\eta'\eta'' x\Big]\nn\\
    &=-\kappa\hat x\oint\frac{1}{f'}\Big[(1+\eta\eta')\mathcal{L}-\frac12\text{Sch}(f)-\frac32\eta'\eta''-\frac12\eta\eta'''\Big]
\end{align}
 where $\text{Sch}(f)=\frac{f'''}{f'}-\frac32\frac{f''^2}{f'^2}$ is the Schwarzian derivative. We can also read the on-shell action by substituting $\psi$ from \eqref{spinorCas} into \eqref{bondryactionJT},
\begin{align}
    I^{\text{\tiny{JT}}}_{\tiny{\text{on-shell}}}=-\hat x\kappa{C}\oint f'=-\hat x\kappa{C}\beta\,.
\end{align}
\subsection{Cangemi-Jackiw supergravity}
Now we apply the same method to the case of $\widehat{\text{CGHS}}$ supergravity with the $A$ field configuration given in \eqref{bndrcd} and the $B$-field being a fundamental field in the Maxwell superalgebra \eqref{superalg}
\begin{align}\label{dilaton}
    B=x^+(u)P_++\dilx(u)P_-+Y(u)J+\dily(u)Z+\phi(u)Q_1+\dino(u) U_2\,.
\end{align}
The inner product of the $B$-field \eqref{dilaton} and variation of the gauge field \eqref{bndrcd} then gives
 \begin{align}\label{variaitonI}
\langle B,\delta A_u\rangle=\dilx\delta\mathcal T+(b\,Y-\dily)\delta\mathcal P+\dino\delta\psi\,.
 \end{align}
 It turns out that $\dily$, $\dilx$ and $\dino$ are independent state-dependent functions and other non-zero elements in \eqref{dilaton} are solved using half of the field equations $\extd B+[A,B]=0$ (along $Z$, $P_{-}$ and $U_2$ generators),
\begin{subequations}\label{eom34}
\begin{align}
     x^+&=\mathcal{T}\dilx+\dily'-\frac{1}{2}\psi \dino \,,\\
     Y&=\mathcal{P} \dilx+\dilx'\,,\\
     \phi&=\psi \dilx-\frac{1}{2}\mathcal{P} \dino-\dino'\,.
 \end{align}
\end{subequations}
 There exist three more equations along $J$, $P_+$  and $Q_1$ which solve the derivatives of dependent fields respectively as
\begin{subequations}\label{eom12}
\begin{align}
     &\qquad Y'=0\,,\\ 
     & \qquad {x^+}'+\mathcal{T}Y-\mathcal{P}x^++\psi\phi=0\,,\\ 
     &\qquad \phi'-\frac{1}{2}\mathcal{P}\phi+\frac{1}{2}\psi Y=0\,.
\end{align} 
\end{subequations}
The last two equations in \eqref{eom12} can be rewritten in terms of independent variables as
\begin{subequations}\label{twoeq}
\begin{align}
    &\mathcal{T}'\dilx+2\mathcal{T}\dilx' -\mathcal{P}\dily'+\dily''-\frac32\psi\dino'-\frac12\psi' \dino=0\,,\\
    &\psi'\dilx+\frac32\psi \dilx'-\dino''+\frac{1}{4}(\mathcal{P}^2-2\mathcal{P}' )\dino=0\,.\label{4.10}
\end{align}
\end{subequations}
Using the field equations \eqref{eom12}-\eqref{twoeq}, we can show that the theory has three independent constants of motion
\begin{subequations}\label{casimirfunct}
\begin{align}
     C_1&\equiv \frac12\left(\mathcal P\dilx+\dilx'\right)\,,\\ 
     C_2&\equiv %x^+\dilx-\frac12\dilyY-\phi\dino=
    \mathcal{T}{\dilx}^2+ {\dily}'\dilx-2\dily C_1-\frac{3}{2}\psi\dino \dilx+\dino'\dino\,,\label{4.17}\\
    C_3&\equiv   \psi{\dilx}^{\frac{3}{2}}-\dilx\left (\frac{\dino}{\sqrt{\dilx}}  \right )' +C_1^2 \frac{\eta}{\sqrt{\dilx}}\,,\label{fermionicconst}
\end{align}
\end{subequations}
where the last constant of motion is fermionic in which we introduced the Grassman variable $\eta$ as
\begin{align}\label{rhoeta}
    \dino {\dilx}^{-\frac{3}{2}}=\left(\frac{\eta}{\sqrt{\dilx}}\right)'\,.
\end{align}

As we will see in a bit, it is very useful to write all free fields in terms of $C_i$ and $(\dily,\dilx)$. As an example, for the $b$-term in \eqref{variaitonI} one can simply verify that 
 \begin{align}
Y\delta\mathcal P&=2C_1\delta\frac{2C_1-\dilx'}{\dilx}=2\delta\frac{C_1^2}{\dilx}+{2}C_1^2\delta\frac{1}{\dilx}+\cdots
\end{align}
where $\cdots$ are total derivative terms (on-shell). Writing $(\mathcal T,\mathcal P, \psi)$ in terms of $C_i$, plugging them in the variation terms of the integrand \eqref{variaitonI} and manipulate the variations such that we have the general pattern in \eqref{Varpatern}, we obtain
\begin{align}
     \dilx \delta \mathcal{T}-\dily\delta\mathcal P&= \delta\frac{C_2}{\dilx}+C_2\delta\frac{1}{\dilx}+2C_1\delta\frac{\dily}{\dilx}+\delta\left ( \frac32\psi\dino+\frac{\dino\dino'}{\dilx} \right )-\delta \dilx\left ( \frac32\frac{\psi\dino}{\dilx}+\frac{\dino\dino'}{\dilx^2} \right )+\cdots \,, \nn\\
 \dino\delta\psi&=\delta \dilx\left ( \frac32\frac{\psi\dino}{\dilx}+\frac{\dino\dino'}{\dilx^2} \right )+\frac12\delta\frac{\dino\dino'}{\dilx}-\delta(\frac{ C_1^2 \dino\eta}{\dilx^2})\nn\\&\qquad\quad-C_1^2\left(-\frac{\delta\dino\eta}{\dilx^2}+\frac{3}{2}\frac{\dino\eta\delta \dilx}{\dilx^3}\right)+\dino \dilx^{-3/2}\delta C_3+\cdots\,.\label{rhodeltpsi}
 \end{align}
   If we use the relation \eqref{rhoeta}, the last term in \eqref{rhodeltpsi} is a total derivative (on-shell) and the next to last term simplifies to
\begin{align}\label{integrabilityC2}
   - \frac12 C_1^2\Big[\delta\left(\frac{\eta\eta'}{\dilx}\right)+\left(\frac{\eta\delta\eta}{\dilx}\right)'\Big]\,.
\end{align}
We are almost done, however, if we compare \eqref{integrabilityC2} with the expected format \eqref{Varpatern} we notice that it is not still of the $C_i\delta V_i+\cdots$ form. We thus use the identity $C_1^2\delta V_1=2C_1\delta (C_1V_1)-\delta (C_1^2 V_1)$ to meet the requirements of \eqref{Varpatern} with $V_1=-\eta\eta'/(2x_1)$ here.\footnote{If we apply this to the $2bC_1^2\delta(1/x_1)$ term, the change in the total variation term and the integrability condition for $x_0$ returns the same term in the final action as it is now.} 

In the Euclidean theory after the Wick rotation of the $u$-coordinate all equations in \eqref{twoeq}-\eqref{integrabilityC2} remain intact with $'\equiv\partial_\tau$ once we also Wick-rotate all independent variables due to their conformal weights as in \eqref{WickRot} and
\begin{align}\label{WickRot2}
    x_0\to x_0\,,\qquad x_1\to -ix_1\,,\qquad\rho\to i^{-1/2}\rho\,,\qquad\eta\to i^{3/2}\eta\,.
\end{align}
% These replacements allow us to work in the Euclidean theory. 
Finally after applying $A_\tau=-i A_u$ and the Wick rotation \eqref{WickRot} and \eqref{WickRot2}, we have 
\begin{align}\label{bondryaction}
    \langle B,\delta A_\tau\rangle&=\delta\left(\frac{C_2}{\dilx}+ \frac32\Big(\psi\dino+\frac{\dino\dino'}{\dilx} \Big)+C_1^2\Big(-\frac{\dino\eta}{\dilx^2}+\frac{2b}{\dilx}+\frac{\eta\eta'}{2x_1}\Big)\right)\nn\\&\qquad+\big(C_2+2bC_1^2\big)\delta\frac{1}{\dilx}+2C_1\delta\Big(\frac{\dily}{\dilx}-C_1\frac{\eta\eta'}{2x_1}\Big)\,.
\end{align}
We thus consider the boundary action as the minus sign of the integral of the first line in the parenthesis in \eqref{bondryaction}.
Upon substituting $C_1$ and $C_2$ from \eqref{casimirfunct} and \eqref{rhoeta} and ignoring boundary of boundary terms we get 
\begin{align}
  I_\partial=-\kappa\oint\left[\mathcal T\dilx-\dily\mathcal P-\frac{\dily\dilx'}{\dilx}+\frac{\dino\dino'}{2\dilx} +\frac{1}{4\dilx}\left(\mathcal P\dilx+\dilx'\right)^2\Big({\frac32}\eta\eta'+2b\Big)\right]\,,
\end{align}
with $\dino\dino'/\dilx=\eta'\eta'' \dilx+\frac12\eta\eta'(\dilx''-\frac12\dilx'^2/\dilx)-\frac12\eta\eta''\dilx'$.
The whole action \eqref{totalaction} is now differentiable   $\delta I\approx0$ if the following integrability conditions are satisfied 
\begin{align}\label{integrMax}
   \delta\oint\frac{1}{\dilx} =0\,,\qquad \delta\oint \Big(\frac{\dily}{\dilx}-C_1\frac{\eta\eta'}{2\dilx}\Big)=0\,.
 \end{align}
It is important to notice that in deriving the integrability conditions \eqref{integrMax} we assumed that $C_1$ and $C_2$ are constants on-shell so that they can be drawn out of the integral. However, the integrability conditions \eqref{integrMax} are now treated completely off-shell.
 We can solve these conditions by introducing two quasi-periodic functions $(f,g)$,
\begin{align}\label{x0x1fg}
    \dilx=\frac{\hat x_1}{f'}\,,\qquad\dily-\frac{1}{4}\eta\eta'(\mathcal{P}x_1+x_1')=\hat x_1\, g'\circ f\,,
\end{align}
where $\circ$ refers to a composition of functions and $g'\circ f=\partial g/\partial f$. The integrability condition \eqref{x0x1fg} signifies that $\hat x_1$ is the zero mode of $x_1$ ($\oint f'=\beta$) where we assume $\delta \hat x_1=0$. Comparing integrability conditions and the ansatz \eqref{integrMax}-\eqref{x0x1fg} with the ones for Jackiw-Teitelboim supergravity \eqref{JTintegrabcond}-\eqref{diffeos}, we find that here the supersymmetric extension appears in the field redefinition of $g'$ rather than $f'$. 
We can rewrite the boundary Lagrangian in terms of ($f,g,\eta$) as fundamental fields by plugging $(\dily,\dilx)$ from \eqref{x0x1fg};
\begin{align}\label{bndryaction23}
   I_\partial[f,g,\eta]&=-\kappa\hat x_1\oint\frac{1}{f'}\Big[\mathcal{T}+\Big(b+\frac{1}{4}\eta\eta'\Big)\frac{\mathcal M}{2}-\Big((g\circ f)'+\frac14\eta\eta''\Big)\mathcal P+(g\circ f)''\nn\\&\qquad\qquad\qquad
    +\eta'\eta''+b\,\text{Sch}(f)+\frac12\eta\eta'''\Big]\,,
\end{align}
where $\mathcal M=\mathcal P^2-2\mathcal P'$. In deriving the final result \eqref{bndryaction23}, the identity $\frac12\oint Af''^2/f'^3=\oint[(A\,\text{Sch}(f)+A'')/f']$ is used  for a given function $A$ and boundary terms are thrown away.

In order to gain more information about the fields $(f,\eta,g)$ we can work out their transformation rule under the large gauge symmetry \eqref{gaugetrans}. Asymptotic symmetries \eqref{gaugetrans} act on the dilaton multiplet $(x_1,x_0,\rho)$ as, 
\begin{equation}
    \begin{split}    \delta x_1&=\varepsilon x_1'-\varepsilon' x_1\,,\\
    \delta x_0&=\varepsilon x_0'-\sigma' x_1+\frac12(\rho\chi)'+\frac12\mathcal{P}\rho\chi\,,\\
    \delta \rho&=-\frac{1}{2}\rho\varepsilon'+\rho'\varepsilon+\frac{1}{2}x_1'\chi-x_1\chi'\,.
\end{split}
\end{equation}From their definition \eqref{x0x1fg}, the following transformation are then induced on $f$, $\eta$ and $g$, 
\begin{subequations}\label{fgtarns}
\begin{align}
  \delta f&=\varepsilon f'\,, \\
  \delta\eta&=\varepsilon\eta'-\frac12\eta\varepsilon'-\chi\,,\\
  \delta (g\circ f)%&= -\sigma+\varepsilon (g\circ f)'+\frac{1}{2\hat{x}_1}(\rho\chi f')+\hat\lambda(f'\eta\chi)\,,\\
  &=-\sigma+\varepsilon (g\circ f)'+\frac12\Big(\eta'+\frac12 \mathcal{P}\eta\Big)\chi\,,
\end{align}
\end{subequations}
where we used \eqref{rhoeta} as $\rho=-\frac12\eta x_1'+\eta' x_1$. The infinitesimal reparametrization transformation $\varepsilon$ in \eqref{fgtarns} confirms that $f$ and $g$ are the diffeomorphisms and a generic (quasi-)periodic function on the circle respectively. Moreover $g$ transforms under $\sigma$ and $\chi$ while $f$ is invariant. The very fact that we can integrate the transformation on the dilaton fields which are generically in terms of $(f',g',\eta')$, to transformation on fields $(f,g,\eta)$ confirms that the field redefinition \eqref{rhoeta} and \eqref{x0x1fg} and thus the integrability conditions \eqref{integrMax} are correct. Of course, it is understood that only a subset of this set of transformations (generated by the subalgebra \eqref{wedge1}) are symmetries of the boundary action \eqref{bndryaction23} which defines an effective theory for the Goldstone modes $(f,\eta,g)$. 

\section{Thermodynamics}\label{secA}
The bosonic part of the $\widehat{\text{CGHS}}$ supergravity \eqref{2.2} admits a Ricci-flat black hole solution which is consistent with the field configuration \eqref{bndrcd} once we apply a finite gauge transformation with the group element $g=\exp(-rP_+)$. The resultant non-zero zwiebein elements $e^+_u=\mathcal T(u)+r\mathcal P(u)$, $e^-_u=1$ and $e^+_r=-1$  lead to the zero-mode solution \cite{Afshar:2019tvp} 
\begin{align}\label{blkhole}
    \extd s^2= 2(\hat{\mathcal P}r+\hat{\mathcal T})\extd u^2-2\extd u\extd r\,.%\qquad X=x_1 r+2x_0
\end{align}
% \textcolor{blue}{if we set $g=\exp(irP_+)$, the metric takes the form\footnote{Where we consider $ds^2=e_E^ae_E^b \delta_{ab}=(dx^1)^2+(dx^2)^2=d(x^1+ix^2)d(x^1-ix^2)=dx_E^+dx_E^-=2\delta_{+-}dx_E^+dx_E^-$, which implies that $\delta_{+-}=2$ where $x_E^\pm=\frac{x^1\pm ix^2}{2}$}
% \begin{align}
%   \extd s^2=2(\hat{\mathcal P}r+\hat{\mathcal T})\extd u^2+2i\extd u\extd r\,.%\qquad X=x_1 r+2x_0   
% \end{align}
% hg
% }
The horizon is at $r_{\text{\tiny{H}}}=-\hat{\mathcal T}/\hat{\mathcal P}$. 
% The black hole solution \eqref{blkhole} thus is the zero mode supersymmetric solution of our BF-theory \eqref{2.2}. 
Using the field equations \eqref{eom34}-\eqref{eom12} for the zero mode solution we have
\begin{align}\label{zeromode}
    \hat Y=\hat{\mathcal P} \hat{x}_1\,,\qquad \hat \phi=\hat \psi \hat x_1\,,\qquad \hat x^+=\hat{\mathcal T} \hat x_1\,,\qquad\hat \dino=0\,,
\end{align}
where all hatted fields are constants. 

Now that we have made the variational principle well-defined (in the Euclidean theory), we are ready to compute the value of the Euclidean on-shell action for this true (black hole) saddle of the action. For simplicity, we consider the zero-mode solution of the theory.  Plugging in the zero-mode solution \eqref{zeromode} after applying the corresponding Wick-rotation \eqref{WickRot} and \eqref{WickRot2} into the first line of \eqref{bondryaction}, the boundary Euclidean on-shell action (the bulk term is zero) is
\begin{align}\label{onshellaction}
    I_{\tiny{\text{on-shell}}}    =-\frac{\kappa\beta}{\hat{x}_1}\left(\hat C_2+\frac{b}{2}\hat Y^2\right)
\end{align}
where $\hat C_2=\hat {\mathcal T}{\hat x}_1^2-\hat{\mathcal P}\hat x_0\hat x_1$.  
\subsection{Holonomy condition}\label{holosec}
 The regularity condition of the black hole solution \eqref{blkhole} at the horizon in the Euclidean signature translates to imposing a Holonomy condition on the Euclidean gauge field $A_\tau=-iA_u$ with $A_u$ being defined in \eqref{bndrcd},
 along the thermal cycle $\tau\sim \tau +\beta$ to be trivial,
 \begin{align}
\text{Hol}(A_\tau)=\exp\left [ \oint A_\tau \right ]\in Z(G)\,.
\end{align}
This condition suggests that the Holonomy belongs to the center of the 2D  Maxwell group $Z(G)$.
The group element $e^{A}$ belongs to the center of the group iff for any arbitrary member of the group $g=e^{X}$ we have $e^A e^X=e^X e^A$ where upon applying the lemma to the Baker–Campbell–Hausdorff formula we must have % where $X=b^+P_++b^-P_-+aJ+a_1Z+\phi Q_1+\gamma U_2$, we have
\begin{align}
 [A,X]+\frac{1}{2}[A,[A,X]]+\frac{1}{3!}[A,[A,[A,X]]]+...=0  \,. 
\end{align}
Imposing this condition to our Euclidean gauge field, with $A=-i\beta\left(\hat{\mathcal T} P_++P_-+\hat{\mathcal P} J+\hat{\psi}Q_1\right)$ and $X$ a generic element in the 2D super-Maxwell algebra,
%=-i \mathcal T P_+^{\text{\tiny E}}-iP_-^{\text{\tiny E}}-\mathcal P_{\text{\tiny}} J_{\text{\tiny E}}-i\psi Q_1$  
% as a gauge field in the 2D super-Maxwell} algebra \eqref{superalg}, 
we have two solutions as the non-trivial center (assuming $\hat{\mathcal P}\neq0$)
\begin{align}\label{holonomycond}
    \hat\psi=0\,,\hat{\mathcal P}=\pm \frac{2\pi n}{\beta}\,,\qquad\text{or}\,\qquad \hat\psi\neq0\,,\hat{\mathcal P}=\pm \frac{4\pi n}{\beta}\,,\qquad n\in\mathbb{N}\,,
\end{align}
and there is no restriction on $\hat{\mathcal T}$ in both cases. As we can see, in the case where the fermionic gauge field $\psi$ is non-zero our solution is regular if $\hat{\mathcal{P}}$ is quantized in $4\pi \times$ temperature $T$. We should emphasize that in \eqref{holonomycond} we should choose the minus sign to ensure that $u$ in the black hole solution \eqref{blkhole} is time-like. Using the value $\hat{\mathcal P}=-4\pi/\beta$ and the on-shell action \eqref{onshellaction} we are now ready to evaluate the entropy of our regular black hole  solution \eqref{blkhole}
\begin{align}\label{entropy}
    S&=-\left(1-\beta\partial_\beta \right)I_{\tiny{\text{on-shell}}}\nn\\
    &=4\pi\kappa\left(r_{\text{\tiny{H}}}\hat{x}_1+\hat{x}_0+4\pi bT\hat{x}_1\right)=4\pi\kappa\left ( \hat{X}_{\tiny{\text{\tiny {H}}}} -b\hat{Y}\right )\,,
\end{align}
where $\hat{X}_{\text{\tiny{H}}}=r_{\text{\tiny{H}}}\hat{x}_1+\hat{x}_0$, is the value of the dilaton field $X$ at the horizon and $\hat Y=-4\pi T\hat x_1$ is the on-shell value of the scalar field $Y$.
This entropy is two times bigger than the entropy for the same black hole solution with zero fermionic gauge component ($\psi=0$). We notice that in this case the value of the zero mode of the auxiliary field, $\hat{Y}$, is also  doubled. The thermodynamic properties of the $\psi=0$ case have been considered in \cite{Afshar:2020dth} where it was shown that the value of the specific heat $C=T\partial_TS$ for the case $b\neq0$ (twisted-$\widehat{\text{CGHS}}$) is finite and linear in $T$. Here for $\psi\neq0$, it will be $C=(4\pi)^2\kappa b T{\hat x}_1$ which is 4 times bigger than the case $\psi=0$.

\paragraph{Supersymmetric solution}
% Above we saw that only a subgroup of asymptotic symmetries preserves the background. Now we can ask 
Is  the supersymmetry preserved by  our  background \eqref{bndrcd} as a `supersymmetric' solution? The requirement that the residual susy transformation does not change our background fields corresponds to $\delta_\chi\mathcal P=\delta_\chi\mathcal T=\delta_\chi \psi=0$. From \eqref{variations} the first requirement is automatically satisfied while other conditions lead to 
\begin{align}
-\frac32\psi\chi'-\frac12\psi'\chi=0\,,\qquad \frac14\hat{\mathcal P}^2\chi-\chi''=0\,.
\end{align}
The solution to these equations (assuming $\hat{\mathcal P}\neq0$) are as follows
\begin{align}\label{KilingSpinor}
    \chi(u)=c_1 e^{\frac{\hat{\mathcal P}u}{2}}+ c_2 e^{-\frac{\hat{\mathcal P}u}{2}}\,,\qquad\psi(u)=c_3\chi(u)^{-3}\,.
\end{align}
Thus the first Holonomy-consistent black-hole solution ($\hat\psi=0$ and $\hat{\mathcal P}=-2\pi n/\beta$) is supersymmetric where the spinor $\chi$ is given by \eqref{KilingSpinor}. In the Euclidean theory,
since we define our boundary theory on a circle $\tau\sim\tau+\beta$ where $\tau=iu$ and $\hat{\mathcal P}\to i \hat{\mathcal P}\equiv\hat{\mathcal{P}}_{\text{\tiny{E}}}=-2i\pi n/\beta$, these solutions are anti-periodic consistent supersymmetric solutions.
% We thus see that only the first Holonomy-consistent case is supersymmetric. These equations have non-trivial periodic solutions in the Euclidean case $u\to iu$ as we will briefly mention below.
% Corresponding to the two Holonomy conditions, we have two situations. the first one is when
% \begin{align}
%     \mathcal{P}=\frac{2\pi n}{\beta}, \quad \psi=0
% \end{align}
% Which leads to,
% \begin{align}
%     \frac14(\frac{2\pi n}{\beta})^2\chi-\chi''=0\,, \qquad \chi=\chi_{_{{ }_{n/2}}}  e^{ \frac{\pi}{\beta}nu}+\chi_{_{{ }_{-n/2}}} e^{- \frac{\pi}{\beta}nu}\,.
% \end{align}
% For the second one we have,
% \begin{align}
%     \mathcal{P}=\frac{4\pi n}{\beta}, \quad \psi\neq 0 
% \end{align}
% Which leads to,
% \begin{align}
%      \frac14(\frac{4\pi n}{\beta})^2\chi-\chi''=0\,, \qquad 
% \chi=\chi_{_{{ }_{n}}}  e^{ \frac{2\pi}{\beta}nu}+\chi_{_{{ }_{-n}}} e^{- \frac{2\pi}{\beta}nu}\,. \\ -\frac32\psi\chi'-\frac12\psi'\chi=0\,,\qquad \psi (u)=\frac{Ye^{\frac{3\pi}{\beta}nu}}{\left ( \chi_{_{{ }_{-n}}}+\chi_{_{{ }_{n}}}e^{ \frac{2\pi}{\beta}nu} \right )^3} 
% \end{align}
\subsection{Stabilizer subalgebra}
% Not all infinite dimensional asymptotic symmetry generators are global isometries of the background. 
The stabilizer of the Wick-rotated ($u\to iu\equiv\tau$) background \eqref{blkhole} can be interpreted as the isometry group of the Euclidean version of the flat connection \eqref{bndrcd}. In order to achieve them we need to solve the stabilizer condition $\delta\mathcal T = \delta\mathcal P=\delta\psi=0$ in the Euclidean case. We thus Wick-rotate the functions and parameters of gauge transformation as mentioned in \eqref{WickRot} such that the form of the variations \eqref{variations} is preserved.

For the zero-mode Euclidean solution consistent with the Holonomy condition ($\hat{\mathcal{P}}_{\text{\tiny{E}}}=-\frac{2\pi in}{\beta}, \psi=0,\mathcal{T}=\hat{\mathcal{T}}$) we are led to a pair of non-trivial differential equations for ($\varepsilon, \sigma, \chi$)
\begin{subequations}
\begin{align}
\delta \mathcal P&=\Big(-\frac{2\pi in}{\beta} \varepsilon+\varepsilon'\Big)'=0\,,\\
\delta \mathcal T&=2\varepsilon'\hat{\mathcal{T}} +\frac{2\pi in}{\beta}\sigma'+\sigma''=0\,,\\
\delta \mathcal \psi&=\frac14\Big(-\frac{2\pi in}{\beta}\Big)^2\chi-\chi''=0\,,
\end{align}
\end{subequations}
which are solved in terms of some integration constants;
\begin{subequations}
\begin{align}
    \varepsilon&=\varepsilon_{{}_0}+\varepsilon_{_{{}_{n}}} e^{\frac{2\pi }{\beta}in\tau}\,,\\ \sigma&=\sigma_{{}_0}+\sigma_{{}_{-n}}{e^{-\frac{ 2\pi}{\beta}in\tau}}-\frac{\hat{\mathcal{T}}\beta}{2\pi i n}\,\varepsilon_{{}_{n}}{e^{\frac{ 2\pi }{\beta}in\tau}}\,,\\ \chi&=\chi_{_{{ }_{n/2}}}  e^{ \frac{\pi}{\beta}in\tau}+\chi_{_{{ }_{-n/2}}} e^{- \frac{\pi}{\beta}in\tau}\,.
\end{align}
\end{subequations}
In this case, the functions $(\varepsilon,\sigma)$ are periodic while the spinor function $\chi$ is anti-periodic. We can thus identify it with the NS sector. In this case, we see that our infinite-dimensional super warped-Virasoro symmetry algebra \eqref{InftyAlg} breaks to the $\mathcal N=1$ graded extension of  Maxwell algebra as was anticipated in \eqref{wedge1} (for $n=1$) as the wedge subalgebra.  

Alternatively, if we consider the second Holonomy-consistent zero-mode solution $(\hat{\mathcal{P}}_{\text{\tiny{E}}}=-\frac{4\pi in}{\beta}, \psi=\hat\psi\neq0,\mathcal{T}=\hat{\mathcal{T}})$ we  have 
\begin{subequations}
\begin{align}
\delta \mathcal P&=\Big(-\frac{4\pi in}{\beta} \varepsilon+\varepsilon'\Big)'=0\,,\\
\delta \mathcal T&=2\varepsilon'\hat{\mathcal T} +\frac{4\pi in}{\beta}\sigma'-\frac32\hat\psi\chi'+\sigma''=0\,,\\
\delta \mathcal \psi&=\frac32\hat\psi\varepsilon'+\frac14\Big(-\frac{4\pi in}{\beta}\Big)^2\chi-\chi''=0\,,
\end{align}
\end{subequations}
which are solved as
\begin{subequations}
    \begin{align}
    \varepsilon&=\varepsilon_{{}_0}
   +\varepsilon_{{}_{2n}}{e^{\frac{4\pi }{\beta}in \tau}}\,,\\ \sigma&=\sigma_{{}_0}+\sigma_{{}_{-2n}}e^{-\frac{4\pi }{\beta}in \tau}{-}\frac{\beta}{4\pi in}\left[\left ( \hat{\mathcal{T}}{-} \frac{3\hat\psi^2\beta}{8\pi i n}\right )\varepsilon_{{}_{2n}}e^{\frac{4\pi }{\beta}in \tau}{-}\hat\psi\left(3\chi_{{}_{-n}}e^{-\frac{2\pi i}{\beta}in u}+\chi_{{}_{n}}e^{\frac{2\pi }{\beta}in \tau}\right)\right],\,\\ \chi&= \chi_{{}_n} e^{\frac{ 2\pi}{\beta}in \tau}+\chi_{{}_{-n}} e^{-\frac{2 \pi}{\beta}in \tau}{+}\frac{\hat\psi \beta}{2\pi in}\varepsilon_{{}_{2n}}\,e^{\frac{4\pi }{\beta}in \tau}\,.
\end{align}
\end{subequations}
In this case, the functions $(\varepsilon,\sigma,\chi)$ are all periodic and we can identify them with the Ramond sector. 

\subsection{Canonical charges}
We can obtain the same value for the entropy in the Lorenzian theory where the Hamiltonian formulation applies and one can define the canonical charges of the model.
  Canonical charges associated with a gauge transformation $A\to A+\extd\epsilon$ in a BF theory like in any other gauge theory are defined on codimension-2 surfaces, which in two dimensions are points. They are defined as \cite{Grumiller:2017qao}
  \begin{align}\label{canonchrg}
      \delta Q[\epsilon]=\kappa\langle\epsilon,\delta B\rangle\,.
  \end{align}
 Here $\epsilon$ is the set of allowed gauge transformations given in \eqref{gaugetrans}. The canonical charge \eqref{canonchrg} associated with the Killing vector $\xi=\partial_u$ which becomes null at the horizon gives the entropy as a conserved quantity from the first law of thermodynamics $T\delta S=\delta Q[\xi]$. A generic gauge transformation in the first order formulation of gravity (the BF-theory in this case) induces a diffeomorphism on-shell with $\epsilon=A_\mu\xi^\mu$. We thus have
\begin{align}
    T\delta S=\kappa\langle A_u,\delta B\rangle\,.
\end{align}
This formula defines the first law of thermodynamics in our BF theory. 
Applying it to the zero mode solution \eqref{zeromode} in the presence of fermionic modes and integrating it (with $\delta T=0$) we recover the entropy \eqref{entropy}. 
\section{Conclusion}
In this paper, we considered the gauge theory formulation of the matterless conformally scaled CGHS supergravity dubbed $\widehat{\text{CGHS}}$ model or CJ gravity. By imposing suitable boundary conditions we addressed different holographic visions on the would-be boundary quantum theory such as asymptotic symmetries, effective action, and entropy. The asymptotic symmetries span the supersymmetric warped Virasoro algebra \eqref{InftyAlg}, where the anti-commutators of the supersymmetric generators are quadratic twisted Sugawara construction of the null $U(1)$ current algebra. The presence of null currents (abelian Heisenberg subalgebra) in the asymptotic symmetry algebra seems to be a generic feature in flat space holography \cite{Barnich:2006av, Barnich:2009se, Bagchi:2010eg, Barnich:2011mi, Barnich:2013axa, Oblak:2016eij, Afshar:2015wjm, Afshar:2013vka, Safari:2019zmc, FarahmandParsa:2018ojt, Afshar:2021qvi}.  Our analysis confirms that this feature remains valid  in the supersymmetric case --- for other studies on asymptotic symmetries in flat space-time, supergravity sees e.g. \cite{Barnich:2014cwa, Fuentealba:2020zkf}. It would be interesting to investigate the unitary and short representation of the infinite-dimensional superalgebra \eqref{InftyAlg}. The boundary action \eqref{bndryaction23} ensures a well-defined variational principle. It is the supersymmetric extension of the warped-Schwarzian theory at level zero \cite{Afshar:2019tvp} and thus should be the geometric action of the corresponding supergroup. 
It would be interesting to derive this action from group-theoretical arguments as the geometric (group) actions on coadjoint orbits of the super warped-Virasoro group see e.g. \cite{Oblak:2016eij, Barnich:2017jgw}. Such extensions of the Schwarzian action are key links between gravity and field theory sides. It is thus curious to obtain this supersymmetric action from the supersymmetric charged SYK model in a double scaling limit in the spirit of \cite{Afshar:2019axx}. Since we have black hole solutions in our theory we computed the entropy by both evaluating  the on-shell action and calculating the canonical charges. The essential difference with the non-supersymmetric case is due to the holonomy condition which necessitates the zero mode of $\mathcal{P}$ which basically is the Rindler acceleration, to be quantized in $4\pi T$ instead of $2\pi T$. This is due to the presence of the fermionic degrees of freedom which double the value of the entropy.

Recently the $\widehat{\text{CGHS}}$ theory is conjectured to acquire a dual quantum matrix ensemble description similar to JT gravity  \cite{Saad:2019lba, Kar:2022sdc, Rosso:2022tsv} which can potentially capture non-perturbative effects. 
Finally, it would be interesting to embed more dilaton-gravity theories in the 2D holography setup --- see recent papers \cite{Grumiller:2021cwg, Grumiller:2020elf, Ecker:2021guy} and their references.
\section*{Acknowledgements}
We thank Blagoje Oblak, Felipe Rosso, and Dmitri Vassilevich for their comments on the draft. We also thank the anonymous referee for his/her comments which led to some improvements in the paper. HA has been supported in part by SarAmadan grants M401432 and M400185. The authors  thank the Iran National Science Foundation (INSF) for supporting this research project (Grant Number: 4000132).

 % \bibliographystyle{fullsort.bst}
 
 % \bibliography{references} 

\providecommand{\href}[2]{#2}\begingroup\raggedright\endgroup

\end{document}